# Fast Calculation of Probabilistic Optimal Power Flow: A Deep Learning Approach

Yan Yang[1], Juan Yu[1], Zhifang Yang[1], Mingxu Xiang[1]
[1]School of Electrical Engineering, Chongqing University, Chongqing, China, 400030
yujuancqu@qq.com

Ren Liu[2]
[2] Dominion Energy, Richmond, VA, United States, 23059

*Abstract*—Probabilistic optimal power flow (POPF) is an important analytical tool to ensure the secure and economic operation of power systems. POPF needs to solve enormous nonlinear and nonconvex optimization problems. The huge computational burden has become the major bottleneck for the practical application. This paper presents a deep learning approach to solve the POPF problem efficiently and accurately. Taking advantage of the deep structure and reconstructive strategy of *stacked denoising auto encoders* (SDAE), a SDAE-based *optimal power flow* (OPF) is developed to extract the high-level nonlinear correlations between the system operating condition and the OPF solution. A training process is designed to learn the feature of POPF. The trained SDAE network can be utilized to conveniently calculate the OPF solution of random samples generated by *Monte-Carlo simulation* (MCS) without the need of optimization. A modified *IEEE* 118-bus power system is simulated to demonstrate the effectiveness of the proposed method.

*Index Terms*-- Probabilistic optimal power flow （POPF）, deep learning, stacked denoising auto encoders (SDAE), Monte-Carlo simulation (MCS).

## I. INTRODUCTION

With high-penetration renewables integrated into power grid, the sharply increasing uncertainty brings about many challenges to power system operation. *Probabilistic optimal power flow* (POPF) has become a powerful and essential analytical tool to tackle uncertainty for power system operation [1]-[2]. Existing solving algorithms for POPF can be generally divided into three types: analytical approaches, approximate approaches, and numerical approaches [3].

Analytical approaches rely on the linearization of non-linear equations in power systems [4]-[5], which has light computational pressure. However, the linearized relationship would be incorrect when input stochastic variables fluctuate greatly, since this kind of method is inapplicable in the complex and highly nonlinear power system.

Approximation approaches are generally faster than the other two categories. The typical methods include *point estimation method* (PEM) [7] and unscented transformation method [8]. However, the accuracy of higher moments cannot be guaranteed. Furthermore, the computational efficiency is seriously affected by a tremendous number of input stochastic variables.

Numerical approaches are based on *Mont-Carlo simulation* (MCS) [9] which is the most accurate probabilistic analysis method. However, it requires enormous sampling and repeatedly solving the nonlinear and nonconvex *optimal power flow* (OPF) problems. The heavy computational burden has become a bottleneck for the practical application of POPF. Hence, many researches have been done for reducing the computation time of the MCS method in POPF.

Generally, improvements for the MCS can be categorized into three aspects: improvement of sampling, speeding up OPF calculation, and parallel acceleration.

- Improved sampling methods are available for maintaining accuracy with fewer samples compared with MCS [1], [10]. The related researches are relatively mature, but the required time is still unattractive for practical application.
- The simplified DC OPF is the most widely used in power industries as the acceleration method of AC OPF [11]. However, it's also limited by the accuracy.
- Parallel calculation becomes an efficient acceleration method with the rapid development of information technology [12]. Unfortunately, it is difficult to spread in power industries because of expensive hardware cost so far.

Regarding existing solving algorithms for POPF, it is difficult to balance the computational speed, accuracy and cost. Fortunately, the fast-developing deep learning techniques provide a promising way to effectively solve the POPF problem. The OPF optimization process can be considered as a nonlinear correlation between the "input" system operating condition and the "output" OPF solution. Deep learning techniques can be used to extract this nonlinear correlation and "predict" the OPF solution for a new system operating condition.

Deep learning techniques allow the development, training, and use of neural networks that are much larger (more layers) than was previously thought possible. Deep neural networks generally can be divided into *convolutional neural networks* (CNNs), *recurrent neural networks* (RNNs) and full-connected networks. CNNs are the go-to methods for any type of prediction problem involving image data as an input. RNNs are designed to work with sequence prediction problems. Full-connected networks are suitable for regression and classification prediction problems. Moreover, Full-connected networks are proved has the ability to approximate any function with high accuracy in theory. Therefore, in this paper, the *stack denoising auto encoders* (SDAE) [13] as an outstanding fully-connected deep neural network is chosen to learn the OPF optimization process. The complex nonlinear characteristics between input data and output data of the OPF problem are extracted through a continuous encoding and decoding process. The SDAE training strategy based on ReLU activation function, RMSProp algorithm, and momentum learning is designed with considering the physical property of the OPF problem. Further, the proposed SDAE-based POPF is capable of improving the computational efficiency of the conventional POPF calculation process by at least 500 times in the modified *IEEE* 118-bus system. Because of the desired computational efficiency and accuracy, the proposed method opens a door to online application of POPF.

This paper is organized as follows. OPF based on SDAE is presented in Section II. POPF based SDAE and MCS is introduced in Section III. Case studies are shown in Section IV, followed by conclusions in Section V.

## II. OPF BASED ON SDAE

The basic structure of SDAE-based OPF is presented in this section. Then, corresponding training strategy for SDAE-based OPF is designed.

### A. Structure of SDAE-based OPF

SDAE-based OPF consists of multiple denoising auto encoders (DAE). Each DAE has four parts: input layer $X$, corruption layer $\tilde{X}$, single middle layer $a$ and output later $Z$. The diagram for the structure of DAE is illustrated in figure 1. $\tilde{X}$ can be obtained by setting a number of $X$ to zero randomly. It is called corruption procedure in (1). By encoding function $e_\theta$ in (2), middle layer $a$ is determined by corrupted input $\tilde{X}$. By decoding function $d_\theta$ in (3), output layer $Z$ of DAE is expected to reconstruct $X$. The detailed computational procedure is as follows:

$$\tilde{X} = C_D(X) \tag{1}$$

$$a = e_\theta(\tilde{X}) = s(W\tilde{X} + b) \tag{2}$$

$$Z = d_\theta(a) = s(W'a + b') \tag{3}$$

where $C_D$ denotes corruption procedure, weight matrix of encoding $W$ is a $p_y \times p_x$ matrix, and biased vector $b$ is a $p_y$ dimensional vector, weight matrix of decoding $W'$ is a $p_x \times p_y$ matrix, $b'$ is a $p_x$ dimensional vector. $p_x$ and $p_y$ are the number of neurons in input layer and output layer, respectively. $s$ is activation function which will be introduced in detail in next section.

SDAE-based OPF is developed by stacking DAE layer by layer, and its structure is illustrated in figure 2. Note that the output layer $Z$ of DAE is not included in the structure of SDAE-based OPF. By continuously encoding the input $X$, the output of SDAE-based OPF $Y_t$ is finally obtained by (4).

$$Y_t = e_\theta^{(n+1)}(e_\theta^{(n)}(...e_\theta^{(1)}(X))) \tag{4}$$

where, $e_\theta^{(l)}$ is the encoding function of $l$th DAE, $l=1, 2, …, n$, and $n$ is the number of DAE in SDAE, $e_\theta^{(n+1)}$ is the encoding function of top layer.

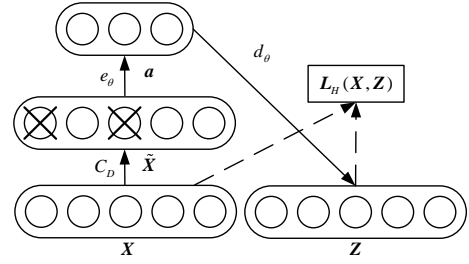

Figure 1. Diagram for the structure of DAE

SDAE exploits the effect of changes in input data on the output data to mine the non-linear features/relationship between them. Therefore, $X$ is designed to only contain the injection active and reactive power of PQ nodes in SDAE-based OPF. Thus, injection power of all renewable nodes and load nodes is the input data. Resistance and reactance of branches and other constant parameters are not selected as input $X$, because they do not vary with system states. Topology is also considered as constant in this paper. In order to meet the requirements of POPF solution, output $Y$ is made up of operation cost, node voltage of all nodes, generator outputs, flow in all branches. Here, SDAE-based OPF is constructed completely.

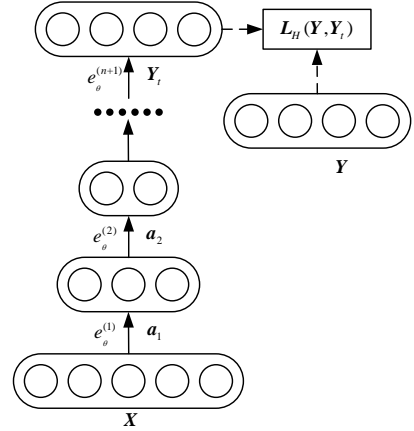

Figure 2. Diagram for the structure of SDAE-based OPF

### B. Training Method for SDAE-based OPF

According to formulas (1) and (3), the neurons of SDAE-based OPF are connected to each other by activation function and encoding parameters $\theta$, namely weight matrix $W$ and biased vector $b$. Therefore, the optimal encoding parameters $\theta$ = {$W$, $b$} are the training objective of SDAE-based OPF.

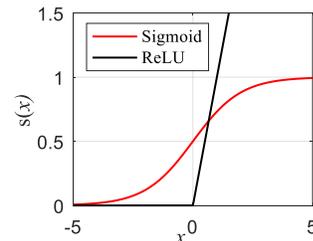

Figure 3. Sigmoid and ReLU activation functions

For activation function, sigmoid function $s=1/(1+e^{-x})$ is widely used in SDAE, because it's plausible biologically [14]. However, when $s$ equals 0 or 1, it will result in gradient vanishing effect due to activation nonlinearity of sigmoid unit. Besides, if the sigmoid function is decided, the OPF solution should be normalized to [0, 1], so that the voltage magnitude and active power of PV node and other constant output can be normalized to 0 or 1. Data close to 0 or 1 is not good for the training as mentioned before. Thus, the *rectified linear unit* (ReLU) function shown in (5) is chosen to make gradient flow well due to the linearity characteristic illustrated in figure 3. What's more, the computation speed of ReLU is also faster than sigmoid function, since there is no need to execute the exponential function and division operations in the neuronal activation process. In order to elude the unbounded behavior of ReLU, all the data is normalized into [0, 1] by min-max normalization method as shown in (6)-(7).

$$s = \max(0, x) \quad (5)$$

$$x = \begin{cases} (x - x_{\min})/(x_{\max} - x_{\min}) & \text{if } x_{\max} \neq x_{\min} \\ x_{\min}/x_{\max} & \text{if } x_{\max} = x_{\min}, x_{\max} \neq 0 \\ x & \text{if } x_{\max} = x_{\min}, x_{\max} = 0 \end{cases} \quad (6)$$

$$y = \begin{cases} (y - y_{\min})/(y_{\max} - y_{\min}) & \text{if } y_{\max} \neq y_{\min} \\ y_{\min}/y_{\max} & \text{if } y_{\max} = y_{\min}, y_{\max} \neq 0 \\ y & \text{if } y_{\max} = y_{\min}, y_{\max} = 0 \end{cases} \quad (7)$$

where $x_{\min}$ and $x_{\max}$ are the minimum and maximum values of the input data in the training samples, respectively. $y_{\min}$ and $y_{\max}$ are the minimum and maximum values of the output data in the training samples, respectively.

For encoding parameters $\theta$, the mean square error function is utilized as a loss function to directly reflect the training effect of SDAE.

$$L_H(y, \hat{y}) = \frac{1}{2}\sum_{k=1}^{d}(\hat{y}_k - y_k)^2 \quad (8)$$

where $y$ is the true value, and $\hat{y}$ is the calculated value by SDAE. $d$ is the dimension of $y$ and $\hat{y}$. As a result, the optimization goal is $\min(L_H(y, \hat{y}))$. SDAE training process contains unsupervised stage and supervised stage. In unsupervised stage, $y$ is the input data of the corresponding DAE. In the supervised stages, $y$ is the output true value of OPF solution. The specific process will be described explicitly in Section III.

In addition, OPF model is nonlinear and nonconvex, which leads to the nonlinear correlation between the input and output of OPF problem. The nonlinear correlation results in deep structure of SDAE-based OPF and a large amount of training samples. *Root mean square propagation* (RMSProp) algorithm decomposes the whole training samples into several batches, and each batch is trained to update parameters $\theta$ in turn. This algorithm can reduce the training pressure of SDAE-based OPF, and relieve the local minimum problem. Moreover, it can adaptively update the learning rate for each connection by keeping a moving average of the squared gradient for each parameter $\theta$. Therefore, RMSProp training algorithm is utilized to obtain optimal $\theta$ of SDAE-based OPF, as shown in (9)-(16).

$$W_{ij}^{(l,T+1)} = W_{ij}^{(l,T)} - \Delta W_{ij}^{(l,T)} \quad (9)$$

$$b_i^{(l,T+1)} = b_i^{(l,T)} - \Delta b_i^{(l,T)} \quad (10)$$

where

$$\Delta W_{ij}^{(l,T)} = \frac{\eta}{\sqrt{\sigma + RR_{ij}^{(l,T)}}} \odot dW_{ij}^{(l,T)} \quad (11)$$

$$RR_{ij}^{(l,T)} = \rho * RR_{ij}^{(l,T-1)} + (1-\rho) * dW_{ij}^{(l,T)} \odot dW_{ij}^{(l,T)} \quad (12)$$

$$dW_{ij}^{(l,T)} = \frac{1}{m}\sum_{k=r}^{r+m}\frac{\partial}{\partial W_{ij}^{(l,T)}} L_H(Y, Y_t) \quad (13)$$

$$\Delta b_{ij}^{(l,T)} = \frac{\eta}{\sqrt{\sigma + R_i^{(l,T)}}} \odot db_i^{(l,T)} \quad (14)$$

$$R_{ij}^{(l,T)} = \rho * R_{ij}^{(l,T-1)} + (1-\rho) * db_i^{(l,T)} \odot db_i^{(l,T)} \quad (15)$$

$$db_{ij}^{(l,T)} = \frac{1}{m}\sum_{k=r}^{r+m}\frac{\partial}{\partial b_i^{(l,T)}} L_H(Y, Y_t) \quad (16)$$

$Y_t$ is OPF solution calculated by SDAE, $W_{i,j}^{(1,T)}$ is the weight from the $j$th neuron in middle layer of DAE in layer $l$-1 to the $i$th neuron middle layer of DAE in layer $l$ after $T$th parameters updating; $b_{i,j}^{(1,T)}$ is the biased value of $i$th neuron of DAE in layer $l$; $\odot$ is Hadamard product; $\eta$ is the learning rate; $r$ is the initial sequence number of samples in mini batch, and $m$ is the sample size of mini batch, $\rho$ is suggested to 0.99, $\varepsilon$ is defaulted as $10^{-8}$.

Finally, in order to further accelerate updating of parameters and surmount local minimum basins of attraction, momentum learning is introduced as additional item. The training process is updated by (17)-(18) from (11) and (14).

$$\Delta W_{ij}^{(l,T)} = p * \Delta W_{ij}^{(l,T)} + (1-p) * \Delta W_{ij}^{(l,T-1)} \quad (17)$$

$$\Delta b_i^{(l,T)} = p * \Delta b_i^{(l,T)} + (1-p) * \Delta b_i^{(l,T-1)} \quad (18)$$

where $p$ is momentum learning rate.

### III. POPF BASED ON SDAE AND MCS

Based on constructed SDAE-based OPF and designed training method, POPF based on SDAE and combined with MCS method is proposed, and the calculation procedures are shown as follows:

*Step 1)* OPF training data acquisition: The training sample for SDAE-based OPF is obtained from experiment and simulation.

*Step 2)* Data preprocessing: This step includes data pre-processing and setting hyper-parameters of SDAE. Here, the input and output of training samples are normalized by min-max normalization method (6)-(7). The training sample input $X$ is corrupted with formula (2). Then, according to the number of training samples, the training samples are divided into $m$ batches. Other hyper parameters of SDAE including number of layers $l$, learning rate $\eta$, momentum learning rate $p$ and number of neurons in each layer are estimated based on the scale of power system.

*Step 3)* Unsupervised training of SDAE-based OPF: according to equations (1)-(5) and (8) construct the loss function $L_H(X, Z)$ of the first DAE with training sample input

$X$. Then, based on applied RMSProp algorithm and momentum learning rate, the parameters updating equations similar to (9)-(18) can be constructed, whereby the optimal parameters for encoder of the first DAE are determined. Afterwards, the output of the middle layer of the first DAE obtained from equation (2) is also the input of the second DAE. Applying the same methods to construct loss function and update parameters, optimal encoding parameters $\theta = \{W, b\}$ for each DAE are determined from bottom to top.

*Step 4)* Supervised training of SDAE-based OPF: According to equations (1)-(8), construct the loss function of SDAE-based OPF $L_H(Y, Y_t)$ with the input and output of training samples. The parameters updating process follows the equations (9)-(18), whereby all the optimal parameters of encoding $\theta = \{W, b\}$ in SDAE are determined. In consequence, SDAE-based OPF is well-trained.

*Step 5)* Sampling of system operating condition: The random variables of system are randomly sampling by MCS. In this paper, the analytical framework of numerical approaches is still maintained. Therefore, improved MCS series methods can also be applied as the sampling method. Besides, the Nataf transformation and the Cholesky decomposition are also allowed to handle the correlations of loads and renewables if needed.

*Step 6)* Probabilistic optimal power flow calculation based on SDAE: All the samples obtained from Step 5) are utilized in the well-trained SDAE-based OPF from Step 4) in a matrix form, then the OPF solutions for all samples can be directly mapped by (4). The final solutions are de-normalized by formulas (6) and (7). On the basis, POPF indexes including mean value, standard variance and probability density can be analyzed effectively.

## IV. CASE STUDIES

In this section, the proposed approach is applied to a modified *IEEE* 118-bus system. Six wind power farms and four photovoltaic stations are connected in the modified system. The detail information of this modified system can be downloaded from [15]. The random characteristic data of wind farm power and photovoltaic station power can be found in [16]. The load in this system follows a normal distribution.

### A. System Information

In order to validate the performance of proposed method, proposed method and other three methods, as shown in Table I, are implemented in the designed test case.

TABLE I. COMPARISON METHODS

| Method | Description |
|---|---|
| M0 | MCS & AC OPF, as reference. |
| M1 | Proposed SDAE-based POPF. There are 3 DAEs in the SDAE, each DAE has 200, 400 and 300 neurons in middle layer, respectively. |
| M2 | PEM [7] |
| M3 | MCS&DC OPF [11] |

The training process is stopped if SDAE-based POPF network meets the condition of early stop method [17] or the number of epochs reaches to the threshold. The left hyper-parameters of SDAE-based POPF are shown in Table II. The number of training samples and validation samples are 50000 and 10000, respectively. All simulations are performed on a 64-bit PC equipped with Intel(R) Core(TM) i7-7500U CPU @ 2.70GHz 32GB RAM in MATLAB environment.

The convergence criterion of MCS is that the variance coefficients of all the probabilistic analysis indexes are less than the given threshold 5% or the number of samples reaches the maximum number of samples 50000.

TABLE II. VALUES OF HYERPARAMETERS

| Training Stage | Hyperparameter | | | |
|---|---|---|---|---|
| | $\eta$ | $M$ | $p$ | epochs |
| Supervised | 0.001 | 500 | 0.9 | 300 |
| Unsupervised | 0.0001 | 500 | 0.9 | 500 |

### B. Analysis of Results

*(i) Accuracy of SDAE-based POPF*

To verify the accuracy of the proposed method, the POPF of case 1 is calculated by methods M0-M3, respectively. The most concerned indexes and the overall calculation accuracy are compared.

Table III shows the most concerned mean and standard deviation of operation cost. It can be seen that the relative errors of the proposed method M1 when calculates mean value and standard deviation is less than 0.4%. That is an attractive accuracy. Unfortunately, the standard deviation calculated by M2 is an imaginary number, which is unreasonable and unavailable. This phenomenon is normal, and it has been discussed in [7]. Besides, the relative errors calculated by M3 are both larger than 5%, which is not a very satisfied value. Therefore, it can be concluded that the proposed method is the most accurate method compared with approximate PEM method M2 and industrial probabilistic DC OPF method M3.

TABEL III
ACCURACY COMPARISON OF OPERATION COST WITH M0-M3 IN CASE 1

| Method | Probabilistic analysis index | | | | |
|---|---|---|---|---|---|
| | Mean | Std | $e_1$(%) | $e_2$(%) | Time(s) |
| M0 | 111463 | 4262 | 0 | 0 | 2882.5 |
| M1 | 111486 | 4248 | 0.02% | 0.33% | 5.0 |
| M2 | 114022 | - | 2.30% | 100% | 11.2 |
| M3 | 105406 | 4618 | 5.43% | 8.35% | 509.2 |

Note: "-" means the index is an unreasonable and unavailable value. $e_1$ and $e_2$ means relative errors of mean value and standard deviation compared with M0, respectively.

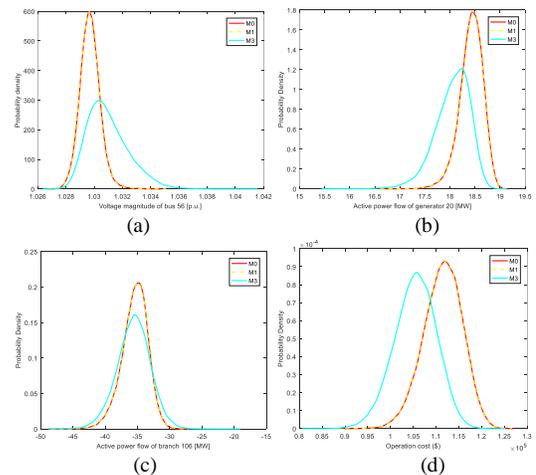

Figure 4. Probability density comparison among M0, M1 and M3 in *Case* 118. (a) Probability density of $V_{56}$. (b) Probability density of $G_{20}$. (c) Probability density of $B_{106}$. (d) Probability density of $f$.

In order to compare the overall calculation accuracy with M0, M1 and M3, the probability that error of bus voltage magnitude exceeds 0.01 p.u. ($P_{ev1}$) and 0.001 p.u. ($P_{ev2}$), probability that error of generator active power output exceeds 3 MW ($P_{eg}$), probability that error of active power of branch exceeds 3MW ($P_{eb}$), and probability that error of operation cost of OPF exceeds 1000 dollars ($P_{ef1}$) and 3000 dollars ($P_{ef2}$) are shown in Table IV, respectively. It can be observed from Table IV that $P_{ev1}$, $P_{eg}$, $P_{ef2}$ and $P_{eg}$ are all approximately zero, which much smaller than that of M3. Therefore, the accuracy of M2 can absolutely meet the basic accuracy requirement in power industries. In addition, this paper tries to focus on higher accuracy, the probability that error of bus voltage magnitude exceeds 0.001 p.u. ($P_{ev1}$) and operation cost of OPF exceeds 1000 dollars ($P_{ef1}$) both less than 0.14%. What's more, figure 4 is illustrated to show the probability density of voltage magnitude of bus 56, active power of generator 20, active power of branch 106, and operation cost. As can be seen figure 4, the proposed method can achieve almost the same calculation accuracy as the accuracy benchmark M0, which is much better than M3.

TABLE IV
CALCULATION ACCURACY FOR OPF BETWEEN M1 AND M3 IN CASE 1

| Method | Probabilistic analysis index | | | | | |
| --- | --- | --- | --- | --- | --- | --- |
| | $P_{ev}$ | $P_{ev2}$ | $P_{eg}$ | $P_{eb}$ | $P_{ef1}$ | $P_{ef2}$ |
| M1 | 0.00% | 0.13% | 0.00% | 0.01% | 0.06% | 0.00% |
| M3 | 1.83% | 48.26% | 23.96% | 47.87% | 91.98% | 76.26% |

*(ii) Speed and generation ability of SDAE-based POPF*

From the calculation time cost in Table III, M0 takes more than 2882 seconds to converge, which illustrates the necessity of fast solving POPF in practical operations. M2 only cost 11.2 seconds to calculate POPF. As mentioned in the introduction, M2 is much faster than the typical numerical methods M0 and M3. It's excited that the proposed approach M1 is faster than M2, and the calculation efficiency is more than 500 times faster than the conventional approach M0.

In conclusion, the proposed approach meets the fast speed and high accuracy requirements for POPF calculation. It opens a door to online application of POPF.

## V. CONCLUSIONS

This paper presents a fast POPF algorithm based on deep learning. Benefiting from the ability of SDAE for extracting the nonlinear future, SDAE-based POPF is proposed. A training strategy based on ReLU activation function, RMSProp algorithm and momentum learning is designed with considering the physical property of the OPF problem. The modified *IEEE* 118-bus test system is simulated to validate the performance of the proposed approach. The simulation results show that the well-trained SDAE-based POPF provides much higher accuracy than the approximate PEM method and the industrial probabilistic DC OPF method. Moreover, combined with the Monte-Carlo method, the proposed method significantly reduces the computation time compared with traditional MCS methods. In the simulation, POPF can be calculated by the proposed approach in second level. Whereby, the proposed approach provides an opportunity for the online application of POPF.

ACKNOWLEDGMENT

The authors are grateful for the support from Chongqing Research Key Program of Basic Research and Frontier Technology (cstc2017jcyjBX0056), Academician-mentoring Scientific and Technological Innovation Project of Chongqing (cstc2018jcyj-yszxX0001), and State Grid Cooperation of China (Research on the efficient calculation based on deep learning for POPF and its application).

REFERENCES

[1] Z. Q. Xie, T. Y. Ji, M. S. Li and Q. H. Wu. "Quasi-Monte Carlo based probabilistic optimal power flow considering the correlation of wind speeds using Copula function," *IEEE Trans. Power Syst*, vol. 33, no. 2, pp. 2239-2247, Mar. 2018.
[2] Yang Y, Yu J, Yang M, Ren P, Yang Z, Wang G. "Probabilistic modeling of renewable energy source based on spark platform with large-scale sample data," *Int Trans Electr. Energ. Syst.* 2018; e2759. https://doi.org/10.1002/etep.2759
[3] Abolfazl Kazemdehdashti, Mohammad Mohammadi, Ali Reaz Seifi, "The generalized Cross-Entropy method in probabilistic optimal power flow," *IEEE Trans. Power Syst*, vol. 33, no. 5, pp. 5738-5748, Mar. 2018
[4] D. Villanueva, J. L. Pazos, and A. Feijoo, "Probabilistic load flowincluding wind power generation," *IEEE Trans. Power Syst*, vo1.26, no.3, pp.1659-1667, 2011.
[5] M. Fan, V. Vittal, G. T. Heydt, and R. Ayyanar, "Probabilistic power flow studies for transmission systems with photovoltaic generation using cumulants," *IEEE Trans. Power Syst.*, vol. 27, no. 4, pp. 2251–2261, Nov. 2012
[6] R.N. Allan, A.M. Leite da Silva, and R.C. Burchett, "Evaluation methods and accuracy in probabilistic load flow solutions," *IEEE Trans. Power App. Syst.*, voLPAS-100, no.5, pp.2539-2546, May 1981.
[7] C. Su, "Probabilistic load flow computing using point estimate method," *IEEE Trans. on Power Syst.,* vol. 20, no. 4, pp. 1843-1851, Nov. 2005.
[8] M. Aien, M. Fotuhi-Firuzabad and F. Aminifar, "Probabilistic Load Flow in Correlated Uncertain Environment Using Unscented Transformation," *IEEE Trans. on Power Syst.*, vol. 27, no. 4, Nov. 2012.
[9] R. Y. Rubinstein and D. P. Kroese, "Simulation and the Monte Carlo method," Hoboken, NJ, USA, Wiley, 2011.
[10] M. Hajian, W. D. Rosehart, and H. Zareipour, "Probabilistic power flow by Monte Carlo simulation with latin supercube sampling," *IEEE Trans. Power Syst.*, vol. 28, no. 2, pp. 1550–1559, May 2013.
[11] M. Ghofrani, A. Arabali, M. Etezadi-Amoli and et al, "Energy Storage Application for Performance Enhancement of Wind Integration," *IEEE Trans. on Power Syst.*, vol. 28, no. 4, pp. 4803-4811, Nov. 2013.
[12] Gan Zhou, Rui Bo, Lungsheng Chien, Xu Zhang, Shengchun Yang, and Dawei Su, "GPU-accelerated algorithm for online probabilistic power flow", *IEEE Trans. on Power Syst.*, vol. 33, no. 1, pp. 1132-1135, Jan. 2013.
[13] Chen Xing, Li Ma, Xiaoquan Yang, "Stacked denoise autoencoder based feature extraction and classification for hyperspectral images," *J. Sensors*, vol. 2016, pp. 1-10, Jan. 2016.
[14] Xavier Glorot, Antoine Bordes, Yoshua Bengio, "Deep Sparse Rectifier Neural Networks," *J. Mach. Learn. Research*, no.15, pp. 315-323, Jan. 2011.
[15] Detailed information of renewables [Online]. Available: https://figshare.com/articles/Case_1_xlsx/7291769.
[16] Zhilong Qin, Wenyuan Li, Xiaofu Xiong, "Incorporating multiple correlations among wind speeds, photovoltaic powers and bus loads in composite system reliability evaluation," *Appl. Energy*, vol. 110, no. 5, pp. 285-294, Oct. 2013.
[17] Garvesh Raskutti, Martin J. Wainwright, Bin Yu, "Early stopping for non-parametric regression: an optimal data-dependent stopping rule," *Journal of Machine Learning Research*, vol. 15, no. 1, pp. 1318-1325, Jan. 2014.